\documentclass[aps,pra,showpacs,superscriptaddress]{revtex4}%
\usepackage{amsmath,amssymb}

\begin{document}

\title{Hardy's criterion of nonlocality for mixed states}

\author{GianCarlo \surname{Ghirardi}}\email{ghirardi@ts.infn.it}%
\affiliation{Department of Theoretical Physics, University of
  Trieste, Italy}%
\affiliation{Istituto Nazionale di Fisica Nucleare, Sezione di Trieste,
  Italy}%
\affiliation{International Centre for Theoretical Physics ``Abdus Salam,''
  Trieste, Italy}%

\author{Luca \surname{Marinatto}}\email{marinatto@ts.infn.it}%
\affiliation{Department of Theoretical Physics, University of
  Trieste, Italy}%
\affiliation{Istituto Nazionale di Fisica Nucleare, Sezione di Trieste,
  Italy}%

\date{\today}

\begin{abstract}
We generalize Hardy's proof of nonlocality to the case of bipartite mixed statistical
 operators, and we exhibit a necessary condition which has to be satisfied by any given mixed state
 $\sigma$ in order that a  local and realistic hidden
 variable model exists which accounts for the quantum mechanical predictions implied by $\sigma$.
Failure of this condition will imply both the impossibility of any local explanation of certain
 joint probability distributions in terms of hidden variables and the nonseparability of the
 considered mixed statistical operator.
Our result can be also used to determine the maximum amount of noise, arising from imperfect
 experimental implementations of the original Hardy's proof of nonlocality, in presence of
  which it is still possible to
 put into evidence the nonlocal features of certain mixed states.
\end{abstract}


\maketitle
%--------------------------------------------------------------------------
\section{Introduction}

Hardy's proof on nonlocality~\cite{hardy} has been referred to as ``the best version of Bell's
 theorem''~\cite{merm}. Such a proof establishes, by resorting to very simple arguments
 which do not involve the consideration of any violation of Bell-type inequalities~\cite{bell,chsh}, a direct
 incompatibility between any local realistic model for almost any bipartite pure entangled state and
 the quantum mechanical predictions concerning properly chosen observables.
However, the most widely used method to deny the existence of a local realistic model for composite
 states consists in the identification of a Bell's inequality which is violated by the state under
 consideration.
Contrary to the case of  pure states in which any entangled vector
 implies the violation of a precise  Bell's inequality~\cite{gisin}, the question of which
 mixed states do violate a Bell's inequality and, as a consequence, do not admit a description in terms
 of a local hidden variable model is more complicated.
In fact, within this wider scenario, there exist non-separable mixed states (that is, statistical
 operators which cannot be expressed as a convex sum of product states) which nonetheless admit a local realistic
 description and do not violate any Bell's inequality~\cite{werner,pop}.
In brief, for mixed states the occurrence of entanglement does not in general rule out a local deterministic
description. Equivalently, no general procedure is known to ascertain
 whether a statistical operator leads to the violation of at least a Bell's
 inequality~\cite{peres,horo,wolf,coll}.

In this paper we exhibit an alternative argument not resorting to Bell's inequalities
 to reject the possibility of  a local realistic description for certain mixed states.
The argument is based on a reformulation and a generalization~\cite{gm} of Hardy's proof of
nonlocality~\cite{hardy}, leading, via simple set theoretic arguments, to an algebraic inequality
 whose violation by a certain mixed state implies the impossibility of a local hidden variable
 model for it. The novelty of the proof derives, on one side, from the fact that it applies to a large
 class of mixed states, contrary to Hardy's proof which was restricted to pure states only, and, on the other side,
 that it holds for bipartite systems whose constituents belong to Hilbert spaces of arbitrary dimensions.
The proof involves both the consideration of the trace distance of the
 considered mixed state from pure Hardy's states (that is, pure entangled states having at least two different
 weights in their Schmidt decomposition) and of the probability for precise joint measurement
 outcomes.
The idea underlying our method is simply that in the vicinity (with respect to the topology induced
 by the trace distance) of a Hardy's state there exist uncountable many (non-separable) mixed states
 which do not admit any local realistic model.
The usefulness of this new nonlocality argument is twofold. First, the possibility of deciding whether
 a given mixed state does exhibit genuine nonlocal features which cannot be reproduced by local classical
 models is extremely important for the theory of bipartite entangled mixed states. In fact, when this
 occurs one can implement efficient quantum communication protocols which cannot be locally reproduced
 by any classical mean~\cite{all}.
Second, our result gives clear indications about the amount of noise which can be tolerated
 when performing an experimental check of nonlocality along the lines indicated by Hardy~\cite{demartini,hardy2}.
More precisely, we can estimate the amount of noise (of the most general kind) affecting the preparation of a
 pure Hardy's state so as to give a mixed statistical operator which still exhibits nonlocal features.

%---------------------------------------------------------

\section{The generalized Hardy's argument}

Given a bipartite state vector $\vert \psi \rangle \in {\mathbb{C}}^{d_{1}} \otimes {\mathbb{C}}^{d_{2}}$,
 where $d_{1}$ and $d_{2}$ are arbitrary positive integers, let us consider its
 Schmidt decomposition in terms of appropriate orthonormal sets of states
 $\left\{ \vert \alpha_{i}(1) \rangle \right\}$ belonging to ${\mathbb{C}}^{d_{1}}$
 and $\left\{ \vert \beta_{i}(2) \rangle \right\}$ belonging to ${\mathbb{C}}^{d_{2}}$,
 respectively,
\begin{equation}
 \label{eq1} \vert \psi(1,2)\rangle = \sum_{i=1}^{\leq min(d_{1},d_{2})}
  p_{i}\vert \alpha_{i}(1)\rangle \otimes \vert \beta_{i}(2) \rangle,
\end{equation}
where the weights $p_{i}$ are positive real numbers satisfying the normalization condition
 $\sum_{i}p_{i}^{2}=1$. Suppose that there exist at least two such weights which
 are different from each other, e.g., $p_{1}\neq p_{2}$, and let
 us denote any state displaying this property as a ``Hardy state".
This is the only hypothesis which, in Hardy's
 proof of nonlocality~\cite{hardy}, is necessary to exhibit a contradiction
 between the existence of a local hidden variable model for a Hardy's state and the
 quantum mechanical predictions for appropriate measurement outcomes.
In our extension of Hardy's proof the mixed states we are going to consider will belong to a neighborhood of a
Hardy's state, whose  size will depend crucially  on the values of $p_{1}$
 and $p_{2}$ of such a state.

To begin with, let us consider an arbitrary bipartite statistical operator $\sigma \in {\cal B}
({\mathbb{C}}^{d_{1}}
 \otimes {\mathbb{C}}^{d_{2}})$ (that is, a positive semidefinite, trace class, trace-one bounded operator)
 whose trace distance from the Hardy state $\vert \psi \rangle$ -- denoted as $D(\sigma, \vert \psi \rangle
 \langle \psi \vert)$ -- is equal to a positive number $\varepsilon$:
\begin{equation}
 \label{eq2}
 D(\sigma, \vert \psi \rangle \langle \psi \vert)\equiv \frac{1}{2}
 Tr\vert \sigma - \vert \psi \rangle \langle \psi \vert \:\vert = \varepsilon,
\end{equation}
where $\vert A \vert \equiv \sqrt{A^{\dagger}A}$ is the positive square root of $A^{\dagger}A$.
 The trace distance between two arbitrary statistical operators $\sigma_{1}$ and $\sigma_{2}$
 represents a good measure to quantify how close are the probability distributions of any measurement
 outcome associated to the two quantum states.
In fact, a well known property of the trace distance is that
\begin{equation}
\label{eq3} \vert Tr[P\sigma_{1}] - Tr[P\sigma_{2}]\vert \leq D(\sigma_{1},\sigma_{2})
\end{equation}
for any projection operator $P$, the expression $Tr[P\sigma_{i}]$ representing the probability
 for the occurrence of a certain measurement outcome when the system is associated with the state
 $\sigma_{i}$.
Since we already know that for any state vector $\vert \psi \rangle$, whose Schmidt decomposition
 involves at least two different Schmidt coefficients, a Hardy's proof of nonlocality
 can be exhibited, our idea is that of trying to determine a neighborhood of $\vert \psi
 \rangle$ [measured in terms of the trace distance $D(\sigma, \vert \psi \rangle
 \langle \psi \vert)=\varepsilon $] such that all mixed states $\sigma$ belonging to it will exhibit
 nonlocal features.
In this way, we will identify  a whole class of bipartite mixed states, belonging to an
 (arbitrary) finite dimensional Hilbert space, which do not admit a local realistic description and
 which are, as a consequence, nonseparable.

To achieve this goal, let us recall the basic steps of  Hardy's argument~\cite{hardy} by resorting
 to the notation we used in our reformulation of that argument~\cite{gm}.
First of all, we define the following two $2\times 2$ unitary matrices $U$ and $V$ whose entries
 depend on the weights $p_{1}$ and $p_{2}$:
\begin{equation}
\label{eq4} U=\frac{1}{\sqrt{p_{1}+p_{2}}}
\begin{bmatrix}
\sqrt{p_{2}} & -i\sqrt{p_{1}} \\
-i \sqrt{p_{1}} & \sqrt{p_{2}}
\end{bmatrix}
\hspace{1.5cm} V=\frac{1}{\sqrt{p_{1}^{2}+p_{2}^{2}-p_{1}p_{2}}}
\begin{bmatrix}
-i(p_{2}-p_{1}) & \sqrt{p_{1}p_{2}} \\
\sqrt{p_{1}p_{2}} & -i(p_{2}-p_{1})
\end{bmatrix}\:.
\end{equation}
Subsequently we consider two orthonormal bases $\left\{ \vert x_{+}(1) \rangle, \vert x_{-}(1)\rangle \right\}$
 and $\left\{ \vert y_{+}(1) \rangle, \vert y_{-}(1)\rangle \right\}$ belonging to the
 two-dimensional linear manifold of the first subsystem spanned by the vectors
 $\left\{ \vert \alpha_{1}(1)\rangle,\vert \alpha_{2}(1)\rangle\right\}$, and
 two bases $\left\{ \vert x_{+}(2) \rangle, \vert x_{-}(2)\rangle \right\}$ and
 $\left\{ \vert y_{+}(2)\rangle, \vert y_{-}(2)\rangle \right\}$ for the two-dimensional
 linear manifold of the second subsystem spanned by the vectors
 $\left\{ \vert \beta_{1}(2)\rangle,\vert \beta_{2}(2)\rangle\right\}$, according to:
\begin{equation}
\label{eq5}
\begin{bmatrix} \vert x_{+}(1) \rangle \\  \vert x_{-}(1) \rangle \end{bmatrix}
=U \begin{bmatrix} \vert \alpha_{1}(1) \rangle \\  \vert \alpha_{2}(1) \rangle
\end{bmatrix} \hspace{1cm}
\begin{bmatrix} \vert y_{+}(1) \rangle \\  \vert y_{-}(1) \rangle \end{bmatrix}
=VU \begin{bmatrix} \vert \alpha_{1}(1) \rangle \\  \vert \alpha_{2}(1)
   \rangle
\end{bmatrix} \hspace{0.1cm}
\end{equation}
\begin{equation}
\label{eq6}
\begin{bmatrix} \vert x_{+}(2) \rangle \\  \vert x_{-}(2) \rangle \end{bmatrix}
=U \begin{bmatrix} \vert \beta_{1}(2) \rangle \\  \vert \beta_{2}(2) \rangle
\end{bmatrix} \hspace{1cm}
\begin{bmatrix} \vert y_{+}(2) \rangle \\  \vert y_{-}(2) \rangle \end{bmatrix}
=VU \begin{bmatrix} \vert \beta_{1}(2) \rangle \\  \vert \beta_{2}(2) \rangle
\end{bmatrix} \:.
\end{equation}
The state $\vert \psi \rangle$ of Eq.~(\ref{eq1}) can then be expressed in three equivalent
 forms in terms of the basis vectors defined in Eqs.~(\ref{eq5}-\ref{eq6}), as:
\begin{eqnarray}
\label{eq7}
   \vert \psi\rangle & = & i\sqrt{p_{1}p_{2}}\,[\,\vert x_{+}(1)\rangle
   \vert x_{-}(2) \rangle + \vert x_{-}(1)\rangle  \vert x_{+}(2) \rangle\,]
   +(p_{2}-p_{1}) \vert x_{-}(1)\rangle \vert x_{-}(2) \rangle
   + \sum_{i>2} p_{i}\vert \alpha_{i}(1)\rangle \vert \beta_{i}(2) \rangle
   \nonumber\\
   & = & i\sqrt{p_{1}^{2}+p_{2}^{2}-p_{1}p_{2}}\, \vert y_{-}(1)\rangle
   \vert x_{-}(2) \rangle +i \sqrt{p_{1}p_{2}}\,\vert x_{-}(1)\rangle \vert
   x_{+}(2) \rangle + \sum_{i>2} p_{i}\vert \alpha_{i}(1)\rangle \vert
   \beta_{i}(2) \rangle \nonumber\\
   & = & i\sqrt{p_{1}p_{2}}\,\vert x_{+}(1)\rangle
   \vert x_{-}(2) \rangle + i\sqrt{p_{1}^{2}+p_{2}^{2}
   -p_{1}p_{2}} \vert x_{-}(1)\rangle \vert
   y_{-}(2)  \rangle+ \sum_{i>2} p_{i}\vert \alpha_{i}(1)\rangle \vert
   \beta_{i}(2) \rangle\:.
\end{eqnarray}
With the aim of displaying the particular set of joint probability distributions
 which conflict with any local hidden variable
 model, we consider the four operators $X_{1},Y_{1},X_{2}$, and $Y_{2}$ having as eigenstates associated to the
 eigenvalues $+1$ and
 $-1$ the orthonormal vectors $ \left\{ \vert x_{+}(1) \rangle, \vert x_{-}(1)\rangle \right\}$,
 $\left\{ \vert y_{+}(1) \rangle, \vert y_{-}(1)\rangle \right\}$, $\left\{ \vert x_{+}(2) \rangle, \vert
 x_{-}(2)\rangle \right\}$, and $\left\{ \vert y_{+}(2) \rangle, \vert y_{-}(2)\rangle \right\}$,
 respectively, while they act as the null operator in the manifolds
 orthogonal to the bidimensional ones corresponding to the nonzero eigenvalues.
According to Eq.~(\ref{eq7}) the quantum joint probabilities concerning the set of
 observables $X_{1},Y_{1},X_{2}$ and $Y_{2}$, when the system is in the state $\vert\psi\rangle$, satisfy the
 following relations:
\begin{eqnarray}
 \label{eq9.1}
 P_{\psi}(X_{1}=+1, X_{2}=+1) &=& 0,\\
 \label{eq9.2}
 P_{\psi}(Y_{1}=+1, X_{2}=-1) &=& 0, \\
 \label{eq9.3}
 P_{\psi}(X_{1}=-1, Y_{2}=+1) &=& 0, \\
 \label{eq9.4}
 P_{\psi}(Y_{1}=+1, X_{2}=0) &=& 0,\\
 \label{eq9.5}
 P_{\psi}(X_{1}=0, Y_{2}=+1) &=& 0\:, \\
 \label{eq9.6} P_{\psi}(Y_{1}=+1, Y_{2}=+1) &=&
  \frac{p_{1}^{2}p_{2}^{2}(p_{1}-p_{2})^2}{(p_{1}^{2}+p_{2}^{2}-p_{1}p_{2})^{2}} \equiv a.
\end{eqnarray}

Since by hypothesis, the weights $p_{1}$ and $p_{2}$ are strictly positive and
 different from each other [thus implying that the parameter $a$ we have defined in Eq.~(\ref{eq9.6})
 is strictly positive], one is able to set up a Hardy-like proof of nonlocality~\cite{hardy} by resorting~\cite{gm}
 to a set theoretic argument leading to a contradiction between the
 considered probability distributions of Eqs.~(\ref{eq9.1})-(\ref{eq9.6}) and the possibility of accounting
 for them by means of a local realistic model where additional hidden variables predetermine the
 outcomes of any conceivable measurement.

In order to generalize such a result to mixed states, let us consider an arbitrary statistical operator
 $\sigma$ having trace distance from  $\vert\psi\rangle\langle\psi\vert$, as defined in Eq.~(\ref{eq2}), equal
 to $\varepsilon >0$.
As a consequence of Eq.~(\ref{eq3}), which gives an upper bound to the difference
 of the probability distributions associated to the different quantum states $\sigma_{1}=\vert \psi \rangle
 \langle \psi \vert $ and $\sigma_{2}=\sigma$, respectively, and taking into account Eqs.~(\ref{eq9.1}-\ref{eq9.6}),
 one obtains:
\begin{eqnarray}
 \label{eq10.1}
 P_{\sigma}(X_{1}=+1, X_{2}=+1) \leq \varepsilon,\\
 \label{eq10.2}
 P_{\sigma}(Y_{1}=+1, X_{2}=-1) \leq \varepsilon ,\\
 \label{eq10.3}
 P_{\sigma}(X_{1}=-1, Y_{2}=+1) \leq \varepsilon, \\
 \label{eq10.4}
 P_{\sigma}(Y_{1}=+1, X_{2}=0) \leq \varepsilon ,\\
 \label{eq10.5}
 P_{\sigma}(X_{1}=0, Y_{2}=+1) \leq \varepsilon ,\\
 \label{eq10.6}
 P_{\sigma}(Y_{1}=+1, Y_{2}=+1) \in [a-\varepsilon, a+\varepsilon].
\end{eqnarray}
Now, suppose that there exists a more complete description of quantum systems than the one characterized by the
 simple assignment of the statistical operator $\sigma$. This description is called a stochastic
 hidden variable model and it consists of (i) a set $\Lambda$ whose elements $\lambda$
 are called hidden variables; (ii) a normalized probability distribution $\rho :\Lambda
 \rightarrow [0,1]$; (iii) a set of probability
 distributions $P_{\lambda}(A_{1}\!=\! a, B_{2}\!=b\!)$ for the measurement outcomes of any pair of observables
 $A_{1}$ and
 $B_{2}$ associated to the first and to the second subsystem respectively, defined for any value
 $\lambda \in \Lambda$, such that
\begin{equation}
 \label{eq11}
  P_{\sigma}(A_{1}=a,B_{2}=b) = \int_{\Lambda}\,d\lambda\,
  \rho(\lambda) P_{\lambda}(A_{1}=a,B_{2}=b).
\end{equation}
The left hand side of Eq.~(\ref{eq11}) gives simply the quantum probability distributions
  concerning the outcomes $\left\{ a,b\right\}$ for
 the joint measurement of the observables $A_{1}$ and $B_{2}$, when the
 system is associated with the statistical operator $\sigma$.
A deterministic hidden variable model (also known as a realistic model) is a particular instance
 of a stochastic model in which the probabilities $P_{\lambda}$ can take only the values $0$ or $1$.
A hidden variable model is called local~\cite{bell2} if the following factorizability condition
 holds for any conceivable joint probability distribution $P_{\lambda}(A_{1}=a,B_{2}=b)$
 and for any value of the hidden variable $\lambda\in \Lambda$
\begin{equation}
 \label{eq12}
 P_{\lambda}(A_{1}=a,B_{2}=b)= P_{\lambda}(A_{1}=a)P_{\lambda}(B_{2}=b),
 \end{equation}
in all cases in which the measurement processes for the observables $A_{1}$ and $B_{2}$
 occur at spacelike separated locations. The locality condition imposes that no causal influence can exist between
 spacelike separated events.
It is worth noticing that it has been proved~\cite{fine} that deterministic and stochastic hidden variable
 models are completely equivalent  when one imposes to them the locality request.
For this reason, in what follows, we will deal, without any loss of generality, only with local realistic models
reproducing the quantum mechanical predictions for the state $\sigma$ in terms of probability distributions
$P_{\lambda}$ assuming only the values $0$ or $1$. Finally, we will denote as $\mu(\Sigma)$ the measure of any
 subset $\Sigma$ of $\Lambda$ with respect to the weight function $\rho(\lambda)$, i.e.,
 \begin{equation}
 \label{12.99}
 \mu(\Sigma)=\int_{\Sigma}\,d\lambda\, \rho(\lambda).
  \end{equation}
To begin with, let us define the following subsets $A,B,C,D$ of the set $\Lambda$ of the hidden variables:
\begin{eqnarray}
 \label{eq13}
 A &= &\left\{ \:\lambda\in \Lambda \:\vert \:P_{\lambda}(X_{1}=1)=1\right\},\\
 B &= &\left\{ \:\lambda\in \Lambda \:\vert \:P_{\lambda}(X_{2}=1)=1\right\},\\
 C &= &\left\{ \:\lambda\in \Lambda \:\vert \:P_{\lambda}(Y_{1}=1)=1\right\},\\
 D &= &\left\{ \:\lambda\in \Lambda \:\vert \:P_{\lambda}(Y_{2}=1)=1\right\}.
 \end{eqnarray}
 Suppose now that a local and realistic description  exists for the mixed state $\sigma$
 and let us consider   the joint probability distribution
 $P_{\sigma}(X_{1}=1,X_{2}=1)$. With our assumptions, we have
\begin{eqnarray}
\label{eq14}
  P_{\sigma}(X_{1}=1,X_{2}=1) & = & \int_{\Lambda}\,d\lambda\,
  \rho(\lambda) P_{\lambda}(X_{1}=1,X_{2}=1)\nonumber \\
  & = & \int_{\Lambda}\,d\lambda\, \rho(\lambda) P_{\lambda}(X_{1}=1)P_{\lambda}(X_{2}=1)
  = \mu[A\cap B],
\end{eqnarray}
where the second equality is implied by the locality condition Eq.~(\ref{eq12}),
 and the third is a consequence of the fact that the product $P_{\lambda}(X_{1}=1)P_{\lambda}(X_{2}=1)$
 does not vanish only within the subset $A\cap B$, where it takes the value one, so that the whole integral
 gives the measure of such a set.
Finally, by resorting to Eq.~(\ref{eq10.1}), we can conclude that $\varepsilon$ is an upper bound
 for the measure of the subset $A\cap B$, that is, $\mu [A\cap B] \leq \varepsilon$.
The situation becomes slightly more complicated when we consider Eq.~(\ref{eq10.2}) and
 impose that there exists a local and realistic model also for such a probability distribution. In fact, by noticing
 that the only outcomes for the observable $X_{2}$ are $-1,0$ and $+1$
 and, as a consequence, that the relation $P_{\lambda}(X_{2}=-1)+P_{\lambda}(X_{2}=0)+P_{\lambda}(X_{2}=+1)=1$
 holds for any $\lambda \in \Lambda$, we have
\begin{eqnarray}
\label{eq15}
  P_{\sigma}(Y_{1}=1,X_{2}=-1) & = &\int_{\Lambda}\,d\lambda\,
  \rho(\lambda) P_{\lambda}(Y_{1}=1)P_{\lambda}(X_{2}=-1)\\
  & = & \int_{\Lambda}\,d\lambda\, \rho(\lambda) P_{\lambda}(Y_{1}=1)[1-
  P_{\lambda}(X_{2}=1) -P_{\lambda}(X_{2}=0)] \\
 & = & \mu[C] - \mu[ B\cap C] - \int_{\Lambda}\,d\lambda\,P_{\lambda}(Y_{1}=1)
 P_{\lambda}(X_{2}=0)
 \end{eqnarray}
Using Eqs.~(\ref{eq10.2}) and~(\ref{eq10.4}), we obtain an upper bound for the
 difference of the measures of the sets $C$ and $B\cap C$
\begin{equation}
\label{eq16} \mu [C]- \mu[B\cap C] \leq 2\varepsilon \:.
\end{equation}
We can now repeat our argument for  all Eqs.~(\ref{eq10.1})-(\ref{eq10.6})
 obtaining in this way two other relations, $\mu[ D]-\mu [A\cap D]\leq 2\varepsilon$
 and $\mu[ C\cap D]\in [a-\varepsilon, a+\varepsilon]$.
Concluding, the following set of constraints on the measure of the considered subsets of $\Lambda$
 have to be satisfied by any local and realistic model  accounting for the
 quantum mechanical predictions implied by any state $\sigma$ satisfying Eq.~(\ref{eq2}):
\begin{eqnarray}
 \label{eq17.1}
 \mu[A \cap B] &\leq & \varepsilon ,\\
 \label{eq17.2}
 \mu[C] - \mu[ B\cap C] & \leq & 2\varepsilon,\\
\label{eq17.3}
 \mu[D]- \mu[A\cap D] & \leq & 2\varepsilon,\\
\label{eq17.4}
 \mu[ C\cap D] & \in & [a-\varepsilon, a+\varepsilon].
\end{eqnarray}
Up to now, no constraint has been imposed on the two  parameters $\varepsilon$, quantifying the distance
 between an arbitrary mixed state $\sigma$ and the precise pure state $\vert \psi
 \rangle$ of Eq.~(\ref{eq1})  and  $a$, which specifies the non-zero probability of Eq.~(\ref{eq9.6}).
 Now we will show that the very assumption that a local realistic description of the implications of the state
 $\sigma$ is possible, implies a precise relation between such parameters.
As a consequence, all states for which
 such a relation is violated do not admit any local realistic description.

In order to find out the  relation constraining the values of $\varepsilon$ and
 $a$ it is useful to resort to the consideration of the complements $\bar{A},\bar{B},\bar{C}$, and $\bar{D}$  in
 $\Lambda$ of the subsets $A,B,C$, and $D$ (that is,
 $\bar{A} \equiv \Lambda- A, \bar{B} \equiv \Lambda -B, \bar{C} \equiv \Lambda -C$, and
 $\bar{D}\equiv\Lambda- D$).
We begin by taking into account Eq.~(\ref{eq17.1}) and the fact that the measure of the whole set $\Lambda$
equals $1$, thus getting:
\begin{equation}
\label{eq18} \mu[ (\Lambda-A)\cup(\Lambda - B)] = \mu [\Lambda-(A\cap B)].
 = 1 - \mu [ A\cap B] \geq 1 - \varepsilon
\end{equation}

\noindent In the same way, using Eq.~(\ref{eq17.4}), we obtain:
\begin{equation}
\label{eq19} \mu[ (\Lambda-C)\cup(\Lambda - D)] = \mu [\Lambda-(C\cap D)]
 = 1 - \mu [ C\cap D] \in [1-a- \varepsilon,1-a+\varepsilon].
\end{equation}
Finally, by considering Eqs.~(\ref{eq17.2}) and~(\ref{eq17.3}) and resorting to simple set manipulations,
 we find that $\mu[ (\Lambda-B)\cup(\Lambda - C)]-\mu[\Lambda-C] \leq 2\varepsilon$ and
 $\mu[ (\Lambda-A)\cup(\Lambda - D)]-\mu[\Lambda-D] \leq 2\varepsilon$.

The previous relations can also be expressed in terms of the complements of the involved sets, in which case
they take the form
\begin{eqnarray}
 \label{eq20.1}
 \mu[\bar{A} \cup \bar{B}] &\geq & 1-\varepsilon ,\\
 \label{eq20.2}
  \mu[ \bar{B}\cup \bar{C}] -\mu[\bar{C}]& \leq & 2\varepsilon,\\
\label{eq20.3}
 \mu[\bar{A}\cup \bar{D}] - \mu[\bar{D}]& \leq & 2\varepsilon,\\
\label{eq20.4}
 \mu[ \bar{C}\cup \bar{D}] & \in & [1-a-\varepsilon, 1-a+\varepsilon].
\end{eqnarray}
These new equations are more suited for deriving the desired constraint between
 $\varepsilon$ and $a$ since they  involve only the union $\cup$ of subsets.
In order to complete our argument, we have first of all to derive three useful relations {\bf RI-RIII}. \\

\noindent {\bf RI}: {\em $\mu[ \bar{A}] \leq \mu [ \bar{A}\cap \bar{D}] +2\varepsilon$ }.\\
\noindent {\em Proof.} This  is easily proved by noticing that
\begin{equation}
 \label{eq21}
 \mu [\bar{A}\cup \bar{D}]-\mu [\bar{D}]= \mu[\bar{A}]-\mu[\bar{A}\cap \bar{D}]\leq
 2\varepsilon
\end{equation}
due to Eq.~(\ref{eq20.3}). $\hfill\blacksquare$\\

\noindent Just in the same way, taking into account Eq.~(\ref{eq20.2}), we get\\

\noindent {\bf RII}: {\em $\mu[ \bar{B}] \leq \mu [ \bar{B}\cap \bar{C}] +2\varepsilon$ }.\\

\noindent Finally, by resorting to elementary set manipulations, we can easily  prove that\\

\noindent {\bf RIII}: {\em $\mu[ \bar{A}\cap\bar{D}] + \mu[\bar{B}\cap\bar{C}]
 \leq \mu [ \bar{C}\cup \bar{D}] + \mu[\bar{A}\cap \bar{B}\cap \bar{C}\cap \bar{D}]$}.\\

{\em Proof.} It is  obvious  that $\bar{A}\cap \bar{D} \subseteq \bar{D} $ and, similarly,
 that $\bar{B}\cap \bar{C}\subseteq \bar{C} $. As
 a consequence $(\bar{A}\cap \bar{D})\cup (\bar{B}\cap \bar{C})\subseteq (\bar{C}\cup \bar{D})$
 and $\mu[(\bar{A}\cap \bar{D})\cup (\bar{B}\cap \bar{C})]\leq
  \mu[\bar{C}\cup \bar{D}]$. By the properties of any measure defined on sets, we have:
\begin{equation}
\label{eq22} \mu[(\bar{A}\cap \bar{D})\cup (\bar{B}\cap \bar{C})]= \mu [ \bar{A}\cap \bar{D}]
 + \mu[ \bar{B}\cap \bar{C}]-\mu [\bar{A}\cap \bar{B}\cap \bar{C}\cap \bar{D}] \leq
 \mu[\bar{C}\cup \bar{D}]
\end{equation}
from which our conclusion holds. $\hfill\blacksquare$\\

Now we have at our disposal all the  relations we need in order to derive the desired constraint
 between $\varepsilon$ and $a$. In fact,
\begin{eqnarray}
\label{eq23}
 \mu[\bar{A}\cup \bar{B}] & = & \mu [\bar{A}] +\mu [\bar{B} ] - \mu[ \bar{A}\cap \bar{B}]
 \leq 4\varepsilon + \mu[ \bar{A}\cap \bar{D}] +\mu [ \bar{B}\cap \bar{C}]-  \mu[ \bar{A}\cap \bar{B}] \\
  & \leq & 4\varepsilon + \mu[ \bar{C}\cup \bar{D}]+ \mu [\bar{A}\cap \bar{B}\cap \bar{C}\cap \bar{D}]
  - \mu[ \bar{A}\cap \bar{B}]\\
   & \leq & 4\varepsilon + \mu[ \bar{C}\cup \bar{D}],
\end{eqnarray}
where the first majorization is implied by {\bf RI} and {\bf RII}, the second is implied by {\bf RIII} and
finally, the last inequality is a trivial consequence of the fact that
 $\mu [\bar{A}\cap \bar{B}\cap \bar{C}\cap \bar{D}] - \mu[ \bar{A}\cap \bar{B}] \leq 0$ since
 $\bar{A}\cap \bar{B}\cap \bar{C}\cap \bar{D}\subseteq \bar{A}\cap \bar{B}$.
 At this point, using the inequalities~(\ref{eq20.1}) and~(\ref{eq20.4}), we end up with the desired
 relation
\begin{equation}
\label{eq25}
 6\varepsilon - a \geq 0.
\end{equation}
We summarize what we have just proved:\\
\noindent {\bf Theorem}: Let us consider an arbitrary Hardy state $\vert \psi(1,2)\rangle = \sum_{i}
 p_{i}\vert \alpha_{i}(1)\rangle \otimes \vert \beta_{i}(2) \rangle$ having two different (non-zero)
 weights $p_{1}\neq p_{2}$, and a statistical operator $\sigma$ such that its trace distance
 $D(\sigma, \vert \psi\rangle \langle \psi \vert)$ from the state $\vert \psi \rangle \langle
  \psi \vert$ equals $ \varepsilon$.
  Then, if a local and deterministic hidden variable model exists for $\sigma$,
  the inequality $6\varepsilon - a\geq 0$ [where
 $a=\frac{p_{1}^{2}p_{2}^{2}(p_{1}-p_{2})^2}{(p_{1}^{2}+p_{2}^{2}-p_{1}p_{2})^{2}}$] has to satisfied for any
  choice of the Hardy state $\vert \psi \rangle$.\\

\noindent The relevance of this theorem derives from the fact that it guarantees that, given a mixed state
 $\sigma$, if there exists a Hardy state $\vert \psi \rangle$ such that the trace distance
 $D(\sigma, \vert \psi\rangle \langle \psi \vert)$ is strictly less than $\frac{a}{6}$ then no local
 realistic description for $\sigma$ can be given.
As a consequence, any such mixed state $\sigma$ cannot be separable (otherwise
 it would admit a local realistic description): thus, the relation of Eq.~(\ref{eq25})  also represents
 a new separability condition for any bipartite mixed state of the Hilbert space ${\mathbb{C}}^{d_{1}}
 \otimes {\mathbb{C}}^{d_{2}}$.

Before concluding, let us underline two important aspects of our result. First of all,
 our nonlocality proof applies to a large class of (bipartite) mixed state. In fact, the set
 of pure entangled states with at least two different weights, i.e., the Hardy states,
 includes the overwhelming majority of all pure entangled states and we have just proved that
 every mixed state $\sigma$ belonging to an appropriate neighborhood of a Hardy state
 (i.e., one of {\em size} less than $a/6$ in the trace distance, where $a$ depends on the considered
 Hardy state) cannot be described by a local realistic model.
Thus, our method allows us to determine a vast class of mixed states exhibiting nonlocal
 features and, as a by-product, a vast class of non-separable states as well.
Secondly, our result might be relevant from a practical point of view. In fact, when implementing
 a Hardy-experiment~\cite{demartini,hardy2} aiming to put into evidence the truly nonlocal
 features of an entangled Hardy's state,
 one has to face the problem of the unavoidably imperfect preparation of such a state.
In general one will be dealing with mixtures of states rather than with the precise pure state one wants to
 study.
Actually, due to unavoidable couplings with the environment, the Hardy's state $\vert \psi \rangle$
 one wishes to
prepare will usually be corrupted by different kind of noises, thus resulting
 in a mixed state $\sigma$ such as
\begin{equation}
\label{eq26} \sigma = p \vert \psi \rangle \langle \psi \vert +(1-p) \tilde{\sigma},
\end{equation}
where $\tilde{\sigma}$ is the (trace-one) statistical operator describing the noise affecting the pure state
$\vert \psi
 \rangle$. Our previous argument shows that, notwithstanding the nonpure nature of the actual state $\sigma$ one
 is dealing with, there exists an interval of values for the parameter $p$ such that
 one can still put into evidence a contradiction between the quantum
 mechanical predictions and those of any local realistic model.
In fact, for the state of Eq.~(\ref{eq26}) one can repeat our earlier arguments, evaluate the trace
 distance $D(\sigma, \vert \psi \rangle \langle \psi\vert)$ and determine
 the precise interval of values for the parameter $p\in (0,1)$ (which quantifies the amount of noise
 in the state) for which a contradiction with locality condition is still present.

%---------------------------------------------------------------

\section{Conclusions}
We have derived a generalization of the original Hardy's proof of nonlocality which works for a particular
 class of bipartite mixed states.
More precisely, we have exhibited a necessary condition which has to be satisfied whenever a local and
 deterministic hidden variable model for a mixed state exists, the condition being that the state itself has to
 lay outside appropriate neighborhoods of all conceivable (entangled) Hardy's states.
 As a consequence, all states which violate the above condition do not admit any local realistic description,
 this in turn implying that they are not separable.

\section{Acknowledgments}
Work supported in part by Istituto Nazionale di Fisica Nucleare, Sezione di Trieste, Italy.
 We thank Professor R. Horodecki and M. Piani for their useful comments.

%--------------------------------------------------------------------

\end{document}